\documentclass[dvips,11pt]{article}
\usepackage{graphicx}
\newcommand{\doublespacing}{\let\CS=\@currsize\renewcommand{\baselinesstrech}
{2.0}\tiny\CS}

\begin{document}

\textwidth 16cm
\newcommand{\bd}{\begin{document}}
\newcommand{\ed}{\end{document}}
\newcommand{\bc}{\begin{center}}
\newcommand{\ec}{\end{center}}
\newcommand{\vs}{\vspace}

\bc {\huge \bf Study of Classical Mechanical Systems } \ec

%\vs{.2cm}

\bc {\huge \bf with Complex Potentials } \ec

\vs{.3cm}

\bc
{\it {\large A. Sinha}{\footnote {e-mail : anjana23@rediffmail.com}} \\
Department of Applied Mathematics, \\
Calcutta University, \\
92, A.P.C. Road, Kolkata - 700 009 \\ INDIA } \ec

\vs{.1cm}

\bc {\it {\large D. Dutta}{\footnote {e-mail :
{debjit$_-$r@isical.ac.in}}}\\
Physics \& Applied Mathematics Unit, \\
Indian Statistical Institute, \\
203, B.T. Road, Kolkata - 700 108 \\ INDIA } \ec

%\vs{.1cm}

\bc {\it and} \ec

%\vs{.1cm}

\bc
{\it {\large P. Roy}{\footnote {e-mail : pinaki@isical.ac.in}} \\
Physics \& Applied Mathematics Unit, \\
Indian Statistical Institute, \\
203, B.T. Road, Kolkata - 700 108 \\ INDIA } \ec

\vs{.3cm}

%\pagebreak

\begin{abstract}

\noindent We apply the factorization technique developed by Kuru
et. al. [Ann. Phys. {\bf 323} (2008) 413] to study complex
classical systems. As an illustration we apply the technique to
study the classical analogue of the exactly solvable ${\cal{PT}}$
symmetric Scarf II model, which exhibits the interesting
phenomenon of spontaneous breakdown of ${\cal{PT}}$ symmetry at
some critical point. As the parameters are tuned such that energy
switches from real to complex conjugate pairs, the corresponding
classical trajectories display a distinct characteristic feature
--- the closed orbits become open ones.

\vs{.5cm}


\end{abstract}

\pagebreak

\section{Introduction}

Fairly recently Kuru {\it et. al.} obtained exact solutions of the
classical analogues of some exactly solvable one-dimensional
quantum mechanical models in the framework of the factorization
technique \cite{kuru-annals}. However, it must be kept in mind
that this method cannot be applied to all classical systems as not
all quantum mechanical models have exactly solvable classical
analogues. On the other hand there have been several attempts to
see if the spontaneous breakdown of ${\cal{PT}}$ symmetry at an
{\it exceptional point} in a certain class of non Hermitian
${\cal{PT}}$ symmetric quantum mechanical systems gets manifested
in the corresponding classical picture
\cite{bender-JMP40,bender-RPP,bender-JPA39,bender-JPA40,nanayakkara-jpa,nanayakkara-pla,kuru-pla}.
For example, the ${\cal{PT}}$ symmetric model $H = p^2 + x^2
(ix)^\epsilon $, was studied numerically in ref.
\cite{bender-JMP40} to show that the exceptional point occurs at
$\epsilon = 0$ in both the quantum as well as the corresponding
classical systems. In a different work \cite{nanayakkara-jpa} the
classical motion of two 1-dimensional non Hermitian systems was
studied --- viz., the complex Harmonic oscillator and the complex
cubic potential --- and an attempt was made to find a connection
between the reality of the spectrum and the regularity of the
classical trajectories. Motivated by these studies, in this work
we intend to extend the factorization technique of ref.
\cite{kuru-annals} to study complex classical systems, which are
classical analogues of non Hermitian quantum mechanical systems.
In particular, we shall apply the technique to study the classical
analogue of the following exactly solvable quantum mechanical
${\cal{PT}}$ symmetric (complex) Scarf II potential \cite{zafar}
\begin{equation}
    V(x) = \displaystyle - v_1 \ {\rm{sech}}^2 x - \hat{v}_1 \ {\rm{sech}}
    \ x \ \tanh x \qquad ; \qquad v_1 \ {\rm{real}} \ , \ v_1 > 0
\end{equation}
For $\hat{v}_1$ pure imaginary (say $\hat{v}_1 = i v_2 $), the
model belongs to the well known category of ${\cal{PT}}$ symmetric
potentials of the form $V(x) = v_1 V_R (x) + i v_2 V_I (x) $ ,
with $V_R(x)$ even, and $V_I(x)$ odd. It is one of the few exactly
solvable quantum mechanical models which exhibits the interesting
phenomenon of spontaneous breakdown of ${\cal{PT}}$ symmetry at a
critical value of the coupling parameter $v_2$. Additionally, its
classical analogue can be solved exactly by means of the
interesting factorization method developed in ref.
\cite{kuru-annals}. Our aim here is to see if the above system
contains periodic or irregular trajectories, and check if the
spontaneous breakdown of ${\cal{PT}}$ symmetry switching energy
from real values to complex conjugate pairs, manifests itself in
the corresponding classical system as well, in terms of
non-periodicity of orbits or any other feature.

The organization of the paper is as follows. To make the paper
self contained, in Section 2 we briefly discuss the factorization
technique of ref.\cite{kuru-annals} to obtain the classical
trajectories of some exactly solvable one dimensional systems, and
extend the same to complex classical systems. Based on the
equations obtained in Section 2, in Section 3 we obtain the
equations of the classical trajectories, the classical momenta and
the phase-space trajectories of a particle under the influence of
the complex Scarf II potential, and plot the corresponding figures
for different values of the coupling parameter $\hat{v}_1$.
Finally, Section 4 is kept for conclusions and discussions.

\section{Formalism}

We start with the one dimensional classical Hamiltonian (in units
$ \hbar = 2m=1$)
\begin{equation}\label{h}
    H (x,p) = p ^2 + V(x)
\end{equation}
where $x,p$ are the canonical coordinates, $V(x)$ is the
potential, and the Poisson bracket of $x$ and $p$ is given as $
\displaystyle \{ x,p \} = 1$. The equations of motion of the
classical particle are given by the Hamilton's equations
\begin{equation}\label{xp}
    \dot{x} = \displaystyle \frac{\partial H}{\partial p} = 2p
    \qquad \qquad , \qquad \qquad
    \dot{p} = \displaystyle - \frac{\partial H}{\partial x} =
    - V^{\prime}(x)
\end{equation} \\
Thus $ \ \ \ddot{x} = 2 \dot{p} = 2 V^{\prime} (x) \ \ $ which on
integration gives the velocity of the particle as
\begin{equation}\label{v}
    \displaystyle v = \frac{dx}{dt} = \pm 2 \sqrt{ E - V(x)}
\end{equation}
$E$ being the energy of the classical particle. In our discussions
since the particle is moving under the influence of complex
forces, time $t$ is treated as a real variable, and the path
$x(t)$ traced out by the particle as well as its velocity $v$ can
take complex values. The roots of the equation $ E - V(x) = 0$
give the locations of the turning points, while the initial
conditions determine the initial velocity of the particle. It may
be mentioned that by conventional notions the velocity of the
particle can take real values only. So the initial positions of
the particle under a real potential lie on the real axis, in
between the turning points. However, if we analytically continue
into the complex plane, a particle under the influence of a
complex force can move about in the complex plane. Thus any point
in the complex plane may be an initial starting point for the
particle.

Our aim here is to obtain exact analytical solutions of the orbits
traced out by the classical particle. For this we shall mainly
follow the factorization technique of ref. \cite{kuru-annals}. We
assume a factorization of the Hamiltonian $H$ in the form
\begin{equation}\label{h-aa}
    H = A^+ A^- + \gamma (H)
\end{equation}
where unlike in usual quantum mechanical factorizations, $\gamma
(H)$ may depend on $H$, and $A^{\pm}$ (not necessarily complex
conjugate) are taken to be of the form
\begin{equation}\label{aa}
    A^{\pm} = \mp i f(x) p + \sqrt{H} g(x) + \varphi (x) + \phi
    (H)
\end{equation}
Furthermore, the functions $A^{\pm}$ and $H$ are assumed to define
a deformed algebra with Poisson Brackets as follows :
\begin{equation}\label{ha-poisson}
    \begin{array}{lcl}
    \displaystyle \left \{ A^{\pm} , H \right \} &=& \pm i \alpha
    (H) A^{\pm} \\ \\
    \displaystyle \left \{ A^+ , A^- \right \} &=& - i \beta (H)
    \end{array}
\end{equation}
where the auxiliary functions $\alpha (H)$, $\beta (H) $ and $\phi
(H)$ are expressed in terms of the powers of $\sqrt{H}$. In case
the quantum version admits bound states with negative energies,
$\sqrt{H}$ should be replaced by $\sqrt{-H}$. Making use of
equations (\ref{xp}), (\ref{aa}) and (\ref{ha-poisson}) we arrive
at the following expressions :
\begin{equation}\label{f}
    \displaystyle f(x) = \frac{2}{\alpha (H)} \left[ \varphi
    ^{\prime} (x) + g^{\prime} (x) \sqrt{H} \right]
\end{equation}
\begin{equation}\label{f-prime}
    \displaystyle f(x) V^{\prime} (x) - 2 f^{\prime} (x)
    \left[ H-V(x) \right] = \alpha (H) \left\{ g(x) \sqrt{H} +
    \varphi(x) + \phi (H) \right]
\end{equation}
\begin{equation}\label{beta}
\begin{array}{lll}
    \beta &=& \displaystyle 2 \sqrt{H} \left[ f^{\prime} (x) g(x) -
    f(x) g^{\prime} (x) \right] - \frac{1}{\sqrt{H}} g(x)
    \left[ 2 f^{\prime} (x) V(x) + f(x) V^{\prime} (x) \right] \\
    \\
    & & \displaystyle + \ 4 f^{\prime}(x) \frac{\partial \phi (H)
    }{ \partial H} \left[ H-V(x) \right]
    - 2 f(x) \left[ \varphi ^{\prime} (x) + \frac{\partial \phi
    (H)}{\partial H } V^{\prime} (x) \right] \\
\end{array}
\end{equation}
Now we construct two quantities of the form
\begin{equation}\label{qpm}
    Q^{\pm} = \displaystyle A^{\pm} e^{\mp i \alpha(H) t}
\end{equation}
which are time dependent integrals of motion. Nevertheless, their
total time derivative vanishes
\begin{equation}\label{dq-dt}
    \displaystyle \frac{dQ^{\pm}}{dt} = \left\{ Q^{\pm} , H \right
    \} + \frac{\partial Q^{\pm}}{\partial t} = 0
\end{equation}
Thus
\begin{equation}\label{q-a}
    \displaystyle \mid Q^+Q^- \mid \ = \ \mid A^+ A^- \mid
\end{equation}
so that the particular  values of the integrals of motion
$Q^{\pm}$ and those of $ A^{\pm} $ may be respectively denoted by
\begin{equation}\label{q}
    Q^{\pm} = \displaystyle c(E) e^{\pm i \theta _0}
\end{equation}
\begin{equation}\label{a}
    \displaystyle A^{\pm} \ = \ \displaystyle c(E) e^{\pm i \left\{ \theta _0
    + \alpha (H) t \right\} }
\end{equation}
where $\theta _0$ is determined from initial conditions, and
\begin{equation}\label{c}
    c(E) = \displaystyle \sqrt{E-\gamma (H) }
\end{equation}
For $c(E)$ to be real, the expression within the square root sign
must be positive. This condition gives the range of energy values
for the classical particle. The solution of (\ref{q}) gives the
trajectories $x(t)$ and momenta $p(t)$ of the corresponding
classical particle in the complex plane. We illustrate our
formalism with the help of an explicit example in the next
section.

%\pagebreak

%\vspace{1cm}

\section{Classical analogue of Complex Scarf II potential}

The quantum version of the ${\cal{PT}}$-symmetric (Complex) Scarf
II potential displays certain interesting features --- \\
~ (i) ~  its discrete spectrum is real below the ${\cal{PT}}$
threshold, often referred to as the Exceptional (or critical)
point, whereas
above it the energy values occur as complex conjugate pairs \\
(ii) ~  its continuous spectrum displays spectral singularity at
the critical point, where the reflection and transmission
coefficients tend to diverge. \\
For the sake of comparison of classical trajectories with other
numerical studies (below and above the critical point), in this
work we shall restrict ourselves to bound states only, hence,
negative energies ($E<0$). The final expression for $V(x) $ should
be of the form
\begin{equation}\label{v-scarf}
    V(x) = \displaystyle - \gamma _0 \ {\rm{sech}}^2 \ {\frac{\alpha
    _0 x}{2}} + 2 \delta \   {\rm{sech}} \ {\frac{\alpha
    _0 x}{2}} \ \tanh {\frac{\alpha _0 x}{2}}
\end{equation}
The parameter $\delta$ plays a crucial role here --- the potential
in (\ref{v-scarf}) is real for real values of $\delta$,
${\cal{PT}}$ symmetric for pure imaginary values of $\delta$, and
a general complex potential (without any ${\cal{PT}}$ symmetry)
for complex values of $\delta$. Our aim in this section is to find
the exact equations for the classical trajectories and momenta of
a particle moving in the complex plane under such a potential for
different $\delta$, with special emphasis on pure imaginary
values, and to see if there is any change in these classical
trajectories and momenta at the onset of spontaneous breakdown of
${\cal{PT}}$ symmetry. The form of the potential in eq.
(\ref{v-scarf}) above demands that in the expression for $A^{\pm}$
in (\ref{aa}) we take the following forms of the functions $g(x) \
, \ \varphi (x) \ , \ \phi (H) $ :
    $$ g(x) \neq 0 \qquad , \qquad \varphi (x) = 0 \qquad , \qquad
    \phi (H) = \displaystyle \frac{\delta}{\sqrt{-H}} $$
so that $A^{\pm}$ reduce to
\begin{equation}\label{a-scarf}
    A^{\pm} = \mp i f(x) p + \sqrt{-H} g(x) +
    \frac{\delta}{\sqrt{-H}}
\end{equation}
Solving equations (\ref{f}), (\ref{f-prime}) and (\ref{beta})
simultaneously gives
\begin{equation}\label{gf-scarf}
    g(x) = \displaystyle \sinh {\frac{\alpha _0 x}{2}}
    \qquad , \qquad
    f(x) = \displaystyle \cosh {\frac{\alpha _0 x}{2}}
    \qquad , \qquad
    \gamma (H) = \displaystyle - \gamma _0 + \frac{\delta ^2}{H}
\end{equation}
So using equations (\ref{q}) and (\ref{c}), and the expression
\begin{equation}\label{aa-s}
    A^+A^- = \displaystyle H + \gamma _0 - \frac{\delta ^2}{H}
\end{equation}
we obtain the value of $c(E)$ as
\begin{equation}\label{c-scarf}
    c(E) = \displaystyle \sqrt{ E + \gamma _0 - \frac{\delta
    ^2}{E}}
\end{equation}
Equation (\ref{c-scarf}) gives the range of values for energy $E$
as $c(E)$ should be real. It is worth remembering here that we are
considering the energy $E$ to be negative. We investigate
the different cases in some detail below : \\

\vspace{.5cm}

\noindent {\bf Case 1 :} $\delta $ is real; $ \delta = \delta _R $
(say) \\
This gives the real Scarf II potential, dealt with in ref.
\cite{kuru-annals}, with classically allowed values of energy $E$
lying in the range
\begin{equation}\label{e-real}
    \displaystyle \frac{- \gamma _0 - \sqrt{ \gamma _0 ^2 + 4
    \delta _R ^2 }}{2} \ < \ E \ < \ 0
\end{equation}

\vspace{.5cm}

\noindent {\bf Case 2 :} $\delta $ is pure imaginary;
$\delta = i \delta _I $ (say) \\
This is the classical analogue of the famous non Hermitian yet
${\cal{PT}}$ symmetric Scarf II potential, which exhibits the
phenomenon of spontaneous ${\cal{PT}}$ symmetry breaking at the
{\it phase transition} point, and real energy values switch to
complex conjugate pairs. Straightforward algebra shows that for
the corresponding classical motion  $c(E) $ reduces to
\begin{equation}\label{c-PTscarf}
    c(E) = \displaystyle \sqrt{ E + \gamma _0 + \frac{\delta_I
    ^2}{E}}
\end{equation}
Since we are dealing with negative energies, the classically
allowed range for $E$ in this case becomes
\begin{equation}\label{e-PT}
    \displaystyle \frac{- \gamma _0 - \sqrt{ \gamma _0 ^2 - 4
    \delta _I ^2 }}{2} \ < \ E \ < \
    \frac{- \gamma _0 + \sqrt{ \gamma _0 ^2 - 4
    \delta _I ^2 }}{2}
\end{equation}
One can check that the classical Hamiltonian, though complex, is
symmetric under parity-time reversal. Furthermore, for $ \gamma _0
\ \geq \ \mid 2 \delta _I \mid $, $E$ is real, corresponding to
exact or unbroken ${\cal{PT}}$ symmetry. However, for $ \gamma _0
\ < \ \mid 2 \delta _I \mid $, energies turn complex. Thus the
classical system undergoes a phase transition at $ \gamma _0 \ = \
\mid 2 \delta _I \mid $.  The corresponding quantum version, too,
undergoes an abrupt {\it phase transition} from exact ${\cal{PT}}$
to spontaneously broken ${\cal{PT}}$ {\it phase} at some critical
value of the coupling parameter $\delta _I$. Our aim in this work
is to see whether this {\it phase transition} gets manifested in
the trajectories traced out by the classical particle.

\vspace{.5cm}

%\pagebreak

\noindent {\bf Case 3 :} $\delta $ is complex; $\delta =
\delta _R + i \delta _I  $ (say) \\
It is easy to observe that the potential (\ref{v-scarf}) is no
longer ${\cal{PT}}$ symmetric; nor is it $\eta$-pseudo Hermitian,
and the energy spectrum is, in general, complex. Naturally, it
does not arouse our interest; so we shall not pursue this case any
further.

%\vspace{.5cm}

%\pagebreak

\subsection{Classical Trajectories}

To plot the classical orbits, we need to derive the expressions
for the trajectories of the classical particle in the complex
plane, under the influence of the potential (\ref{v-scarf}). For
this purpose we assume the most general form of $c(E)$,
viz., $c(E) = c_R (E) + i c_I (E) $ \\
Putting $\delta = \delta _R + i \delta _I$ in the expression for
$A^{\pm}$, i.e., $ A^{\pm} = \displaystyle c(E) e^{\pm i \left(
\theta _0 + \alpha _0 \sqrt{-E} t \right) } $, we obtain
\begin{equation}\label{xp-scarf}
\begin{array}{lll}
    & & \displaystyle \mp ip \cosh {\frac{\alpha _0 x}{2}} + \sqrt{-E}
    \sinh {\frac{\alpha _0 x}{2}} + \frac{ \delta _R + i \delta
    _I}{\sqrt{-E}} \\ \\
    & & \ = \ \displaystyle c_R (E) \cos \left( \theta _0 +
    \alpha _0 \sqrt{-E} t \right) \mp c_I (E) \sin \left( \theta _0 +
    \alpha _0 \sqrt{-E} t \right) \\ \\
    & & \ \ \ \ \ \ + \ i \left\{ c_I  (E)
    \cos \left( \theta _0 + \alpha _0 \sqrt{-E} t \right)
    \pm c_R(E) \sin \left( \theta _0 + \alpha _0 \sqrt{-E} t \right)
    \right \} \\
\end{array}
\end{equation}
and consequently, the final form of the trajectory $x(t)$ and
momenta $p(t)$ as
\begin{equation}\label{x-scarf}
    x(t) = \displaystyle \frac{2}{\alpha _0}  \sinh ^{-1} \left\{
    \frac{\delta _R - c_R (E) \sqrt {-E} \cos \left( \theta _0 +
    \alpha _0 \sqrt{-E} t \right) \pm
    c_I (E) \sqrt {-E} \sin \left( \theta _0 +
    \alpha _0 \sqrt{-E} t \right)   }{E} \right \}
\end{equation}

\begin{equation}\label{p-scarf}
    p(t) = \displaystyle \frac{ E \ \left\{
    - c_R (E) \sin \left( \theta _0 +
    \alpha _0 \sqrt{-E} t \right) \pm
    c_I (E) \cos \left( \theta _0 +
    \alpha _0 \sqrt{-E} t \right) \pm \frac{\delta _I}{\sqrt{-E}}
    \right \}}{ \left[ E^2 + \left\{
    \delta _R - c_R (E) \sqrt {-E} \cos \left( \theta _0 +
    \alpha _0 \sqrt{-E} t \right) \pm
    c_I (E) \sqrt {-E} \sin \left( \theta _0 +
    \alpha _0 \sqrt{-E} t \right)
    \right\}^2  \right]^{1/2}}
\end{equation}
Equation (\ref{x-scarf}) describes the position and equation
(\ref{p-scarf}) the momentum of a classical particle in the
complex plane, as it is under the influence of complex forces.
$x(t)$ and $p(t)$ given above are complex, and therefore we shall
plot the real and imaginary parts of the position and momentum
separately, to gain an insight into the behaviour of the
particle's motion. In this work we shall primarily concentrate on
purely imaginary values of $\delta = i \delta _I$, to see if any
connection exists between the reality of the spectrum and
regularity of the classical orbits.

\subsection{${\cal{PT}}$ symmetric Scarf II potential :
$ \delta \ = \ i \ \delta _I$ }

Initially we start with bound states with real negative energies,
in the unbroken ${\cal{PT}}$ {\it phase} in the quantum version.
The corresponding classical condition for real energies is $ 2
\mid \delta _I \mid  < \gamma _0 $. To plot the exact trajectories
it is essential to have a knowledge about the initial condition
$x(0)$. The classical turning points are given by the roots of the
equation $E-V(x) = 0$, for which we need the precise value of $E$.
The turning points are obtained in the form $x_{\pm} = \pm a +
ib$, and hence are symmetric with respect to the imaginary axis.
This is expected as the classical turning points of a complex
${\cal{PT}}$-symmetric potential for real energies, $E$,
essentially occur as $(z,-z^*) : (-a + ib, a + ib)$ \cite{zafar}.
We plot the orbits traced out by the classical particle in the
complex plane, in Fig. 1, for parameter values $\alpha _0 = 2 \ ,
\delta = 2i \ , \ \gamma _0 = 6$. Equation (\ref{e-PT}) suggests
that classical energies should lie between $-0.763932$ and
$-5.23607$. We plot the trajectories for $E=-3$. Any other value
of energy within the specified range gives similar results. The
classical turning points for this particular set of parameters are
$ \{ \pm 0.781368 - 0.528945 i \} $, $ \{ \pm 0.781368 - 2.61265 i
\}, $ etc., i.e. of the type $(z_1,-z_1 ^*) \ , \ (z_2,-z_2 ^*) $
, etc.

{\begin{figure}[hp]
\begin{center}
\scalebox{0.6}{\includegraphics{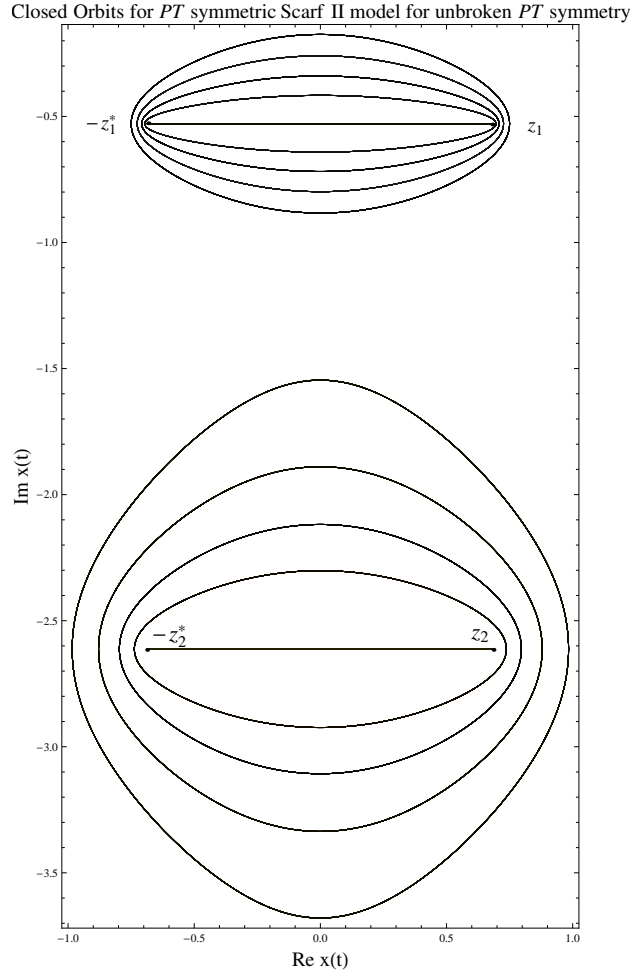}}
\label*{}\caption{\small {The classical trajectories for parameter
values $E=-3 \ , \ \alpha _0 = 2 \ , \ \delta = 2 i \ , \gamma _0
= 6 $ }}
\end{center}
\end{figure}}

\pagebreak

In case the classical particle begins its motion on the real line,
within the turning points or from any one of the turning points
itself, say $z_1$, the particle can only oscillate between the
turning points $(z_1 , - z_1 ^*)$. Similar is the picture for the
particle starting from any position on the real line within the
turning points $(z_2 , - z_2 ^*)$, or from any one of them. These
paths are shown by the horizontal lines in Fig. 1. A trajectory
joining any other pair of turning points (e.g. $z_1, -z_2 ^*$ or
$z_1,z_2$ etc.) is forbidden because the particle is under the
influence of a ${\cal{PT}}$ symmetric potential. However, if the
particle starts its motion from any other point in the complex
plane, the oscillatory trajectories are surrounded by closed
orbits of a definite periodicity. As seen in Fig. 1, the
trajectories are closed curves, which do not cross. The time
period for each of the orbits in Fig. 1 is calculated to be
$3.6276$. Thus our observations for this exactly solvable
analytical model are similar to those obtained in ref.
\cite{bender,bender-JMP40} by numerical methods. In Fig. 2, we
plot the momenta of the classical particle in the complex plane,
and in Fig. 3 we plot the phase space curves, for the same set of
parameter values as in Fig. 1. In these cases, depending on the
initial starting point, the momenta plots as well as the phase
space plots are closed curves without any crossings. However,
while the trajectory plots and the phase space curves are
symmetrical with respect to the imaginary axis, the momenta curves
are symmetric with respect to the real axis.

{\begin{figure}[hp]
\begin{center}
\scalebox{0.7}{\includegraphics{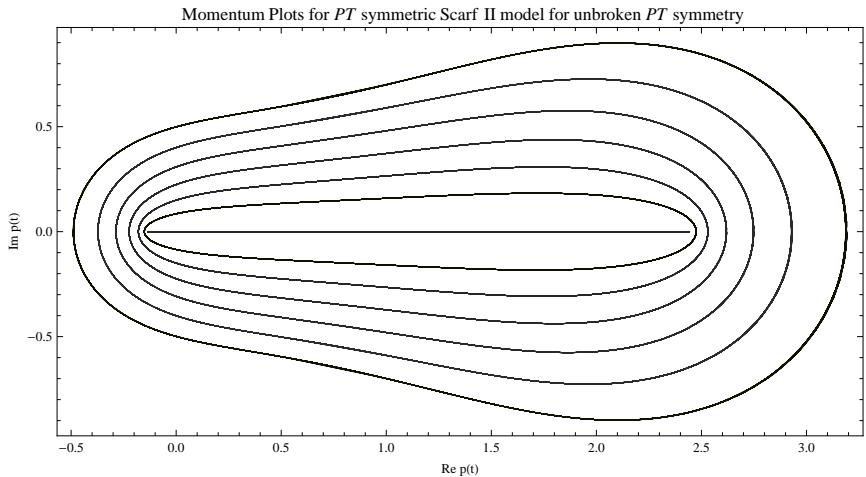}}
\label*{}\caption{\small {The classical momenta for parameter
values $E=-3 \ , \ \alpha _0 = 2 \ , \ \delta = 2 i \ , \gamma _0
= 6 $ }}
\end{center}
\end{figure}}

%\pagebreak

{\begin{figure}[hp]
\begin{center}
\scalebox{0.7}{\includegraphics{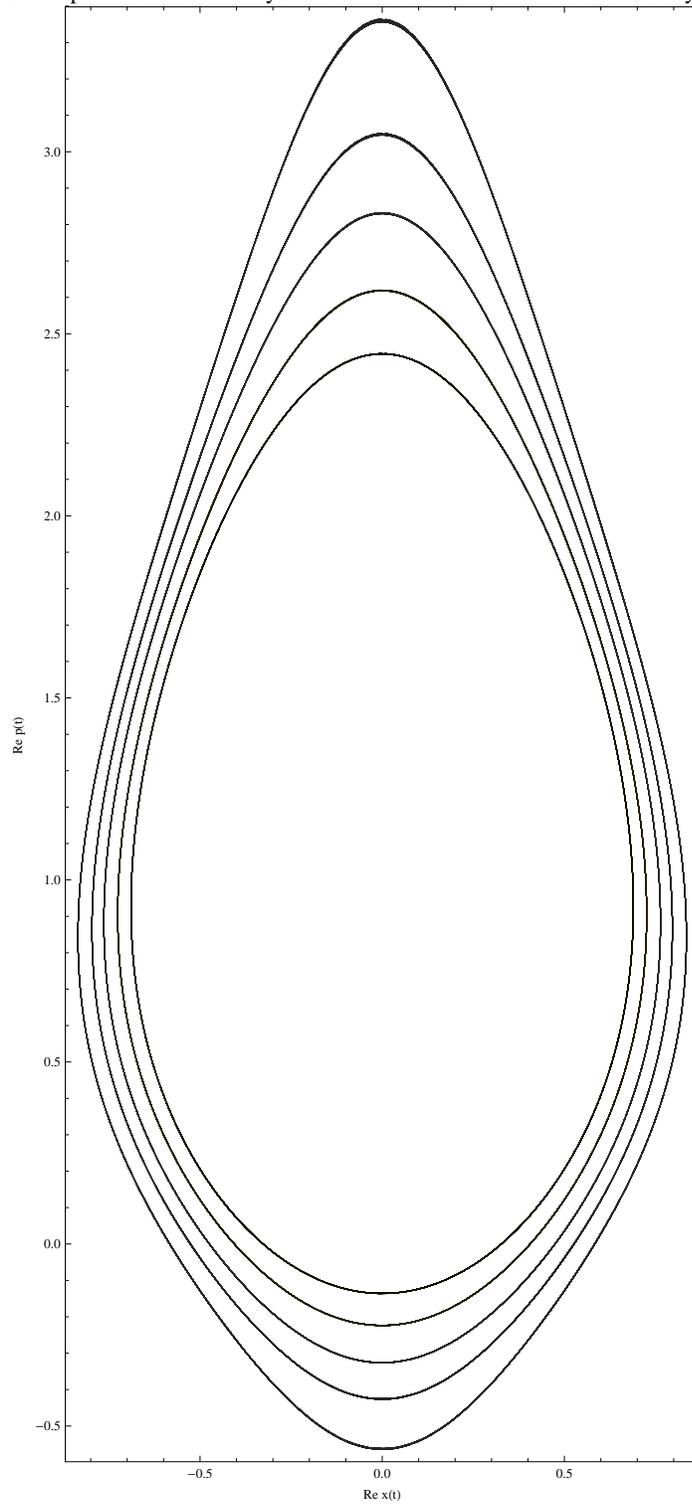}}
\label*{}\caption{\small {The real part of phase space
trajectories for parameter values $E=-3 \ , \ \alpha _0 = 2 \ , \
\delta = 2 i \ , \gamma _0 = 6 $ }}
\end{center}
\end{figure}}

%\vspace{.5cm}

%\pagebreak

Now if the parameter values are such that $ 2 \mid \delta _I \mid
\ > \ \gamma _0$ , then equation (\ref{e-PT}) suggests that the
energies turn out to be complex conjugate pairs, in spite of the
Hamiltonian being ${\cal{PT}}$ symmetric. This is referred to as
the spontaneously broken ${\cal{PT}}$ symmetric {\it phase} in the
quantum version. The classical trajectories are plotted in Fig. 4,
for parameter values $ \ \alpha _0 = 2 \ , \ \delta = 2 i \ ,
\gamma _0 = 3 $. From eq. (\ref{c}), the energies for this set of
parameter values should lie between $ -1.5 + 1.32288 i $ and $
-1.5 - 1.32288 i $. We have taken $E = -1.5 - 0.3i $ for the plots
in Fig. 4. The abrupt {\it phase transition} in the quantum
version gets manifested in the classical motion, too. Irrespective
of the starting point of motion, all of a sudden the closed orbits
become open and the trajectory loses its periodicity. The turning
points are no longer of the type $ (z , - z^* )$, and symmetry is
lost. For this particular set of parameters the turning points
occur at ($ -0.102199 - 0.470998 i \ , \  0.102199 - 2.67059 i \ ,
\ -1.40526 - 1.3489 i \ , \ 1.40526 - 1.79269 i $) etc, i.e. are
of the form $(a-ib_1 \ , \ -a-ib_2 \ , c -id_1 \ , -c -i d_2 )$.
The momenta plots and phase space trajectories, too, lose their
regular behaviour --- the curves are no longer closed ones, nor do
they show any definite periodicity.

{\begin{figure}[hp]
\begin{center}
\scalebox{0.7}{\includegraphics{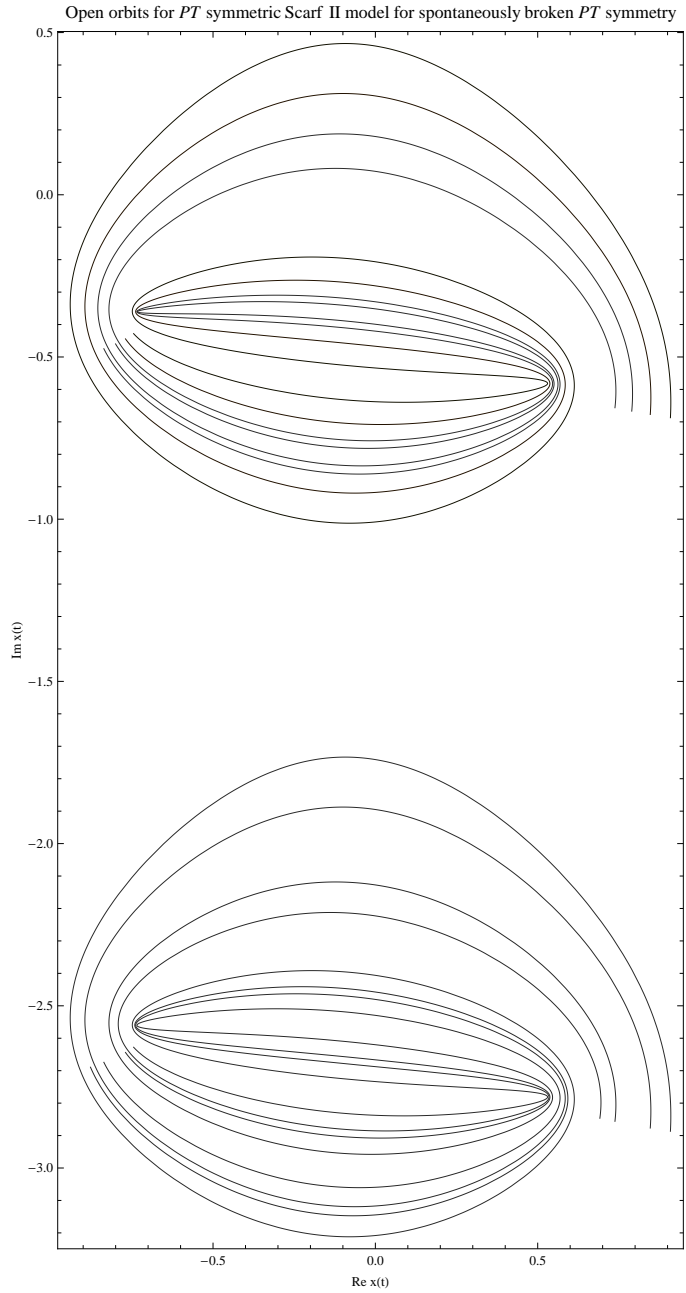}}
\label*{}\caption{\small {The classical trajectories for parameter
values $E=-1.5 - .3i \ , \ \alpha _0 = 2 \ , \ \delta = 2 i \ ,
\gamma _0 = 3 $ }}
\end{center}
\end{figure}}

%\vspace{.5cm}

%\pagebreak

\section{Conclusions}

To conclude, we have extended the factorization technique of ref.
\cite{kuru-annals} to study complex classical systems. As an
illustration we have applied the technique to study the exactly
solvable classical analogue of the exactly solvable, non Hermitian
yet ${\cal{PT}}$ symmetric Scarf II model
    $$ V(x) = \displaystyle - v_1 \ {\rm{sech}}^2 x -
    i \ v_2 \ {\rm{sech}} \ x \ \tanh x \qquad ; \qquad v_1 > 0,
    \ v_1 \ {\rm{real}} $$
Since the particle is under the influence of complex forces, it is
expected to move about in the complex $x$ plane. The quantum
version of this particular Hamiltonian exhibits an abrupt {\it
phase transition} at some critical value of $v_2$, switching
energy values from real to complex conjugate pairs. On the other
hand, for complex values of $v_2$, there is no space-time
reflection symmetry, and energies are always complex. We have
given special emphasis on real values of $v_2$ in this work, so as
to study the effect of spontaneous ${\cal{PT}}$ symmetry breaking
on the classical trajectories and momenta. Employing the
factorization technique of ref. \cite{kuru-annals}, we have found
the equation of the path the classical particle traces out in the
complex $x$ plane, and plotted the corresponding orbits in Fig. 1
and Fig. 4. It is observed that so long as ${\cal{PT}}$ symmetry
is unbroken, the motion of the corresponding classical particle is
bounded, with all trajectories having the same periodicity ---
Fig. 1. Furthermore, the orbits are symmetric with respect to the
imaginary axis, and the classical turning points are of the form
$(z, -z^*)$. The spontaneous breakdown of ${\cal{PT}}$ symmetry
resulting in switching of energy values from real to complex
conjugate pairs, has an interesting manifestation in the
corresponding classical picture
--- the closed periodic orbits abruptly become irregular and open,
as shown in Fig. 4. However, none of the trajectories cross each
other, as is expected for systems symmetric under ${\cal{PT}}$.

We have also plotted the momenta of the classical particle in the
complex plane in Fig. 2, and the phase space trajectories in Fig.
3, so long as ${\cal{PT}}$ symmetry is unbroken and energies are
real. Here, too, the curves are closed, and of definite
periodicity. While the trajectory plots and phase space plots are
symmetric with respect to the imaginary axis, the momenta plots
are symmetric with respect to the real axis. At the onset of
spontaneous breakdown of ${\cal{PT}}$ symmetry at the exceptional
point, each of these plots abruptly loses its regular pattern, and
the closed curves become open.

To summarize, the observed deviation from regularity in the
classical orbits at an exceptional point, for the analytical model
considered in this work, is similar to that shown in earlier works
\cite{bender-JMP40, nanayakkara-jpa}. However, we follow the
factorization approach given in \cite{kuru-annals} and thus our
technique is different from the numerical / perturbative studies
done in ref. \cite{bender-JMP40, nanayakkara-jpa}. Additionally,
the model considered here is exactly solvable, both in its quantum
as well as classical versions.

\section{Acknowledgement}

We thank the unknown referees for their constructive criticisms.
One of the the authors (A.S.) thanks the Dept. of Science and
Technology, Govt. of India, for financial assistance through its
grant SR/WOS-A/PS-06/2008.

\end{document}